\documentclass{aa}
\usepackage{graphicx}
\begin{document}
\title{Modelling the Recurrent Nova CI Aql in Quiescence}

\author{C. Lederle \and S. Kimeswenger}

\offprints{S. Kimeswenger,\\
\email{stefan.kimeswenger@uibk.ac.at}}

\institute{Institut f\"ur Astrophysik der
Leopold--Franzens--Universit\"at Innsbruck, Technikerstr. 25,
A-6020 Innsbruck, Austria (http://astro.uibk.ac.at)}

\date{Received , ; accepted , }

\abstract{We present detailed photometric investigations of the
recurrent nova CI Aql. New data obtained after the 2000 outburst
are used to derive a 3D geometrical model of the system. The
resulting light curves clearly indicate the existence of an
asymmetric spray around the accretion disk, as claimed in the past
e.g. for the super-soft X-ray source CAL87 in the LMC. The
simulated light curves give us the mass transfer rates varying
from $\dot M \approx 2.5 \times 10^{-8}\,\,{\rm M}_\odot\,{\rm
yr}^{-1}$ in 1991-1996 to \mbox{$5.5 \times 10^{-8} < \dot M < 1.5
\times 10^{-7}\,\,{\rm M}_\odot\,{\rm yr}^{-1}$} in 2001/2002. The
distance and the interstellar foreground extinction resulting from
the model are 1.55~kpc and \mbox{E$_{\rm B-V}$ = 0\fm98}
respectively. During fast photometry sequences in 2002 short
timescale variations ($t_{\rm F} \approx 13$ minutes) of the mass
loss are found. Moreover a change in the orbital period of the
system is detectable and results in a mass loss of \mbox{$2.2
\times 10^{-6} < \Delta M < 5.7 \times 10^{-6}\,\,{\rm M}_\odot$}
during the nova explosion.  }

\maketitle
%

\section{Introduction}
CI Aql is one of the 9 to 10 members of the class of recurrent
novae: U~Sco, V394~CrA, RS~Oph, T~CrB, V745~Sco, V3890~Sgr, T~Pyx,
 CI~Aql and IM~Nor (Webbink et al. \cite{We87};
Sekiguchi \cite{Sec}; Schmeja et al. \cite{schmeja}; Liller
\cite{iauc7789}). Webbink et al. (\cite{We87}) also mention
V1017~Sgr as possible class members although its status is still
not clear.
The first known outburst of CI Aql was discovered on Heidelberg
plates recorded in June 1917 (Reinmuth \cite{ci_1}). Williams
(\cite{ci_2}) completed the light curve by using records on
Harvard College Observatory patrol plates. Schaefer (\cite{ci_3})
found another outburst in 1941 again on Harvard plates. Schaefer
argues, it might be a recurrent nova with a timescale of 20 years
and the 1960 and 1980 outburst were missed. As the timescales of
other recurrent novae often change and as there are no
observations available, we assume for our calculations a
quiescence phase of 60 years before the 2000 event. Anyway this
does not affect the results of the model presented here but our
resulting pre--outburst accretion rate indicates a long recurrence
timescale. CI Aql was found to be an eclipsing binary system with
a period of 0\fd618355(9) by Mennickent \& Honeycutt (\cite{M_H}).
It is, to our knowledge, the only eclipsing system investigated
photometrically in such detail before and after an outburst.
Following the classification of Sekiguchi (\cite{Sec}), CI\,\,Aql
is of U\,\,Sco subclass (slightly evolved main sequence star and
accreting white dwarf). For a detailed discussion of the outburst
data we refer to Matsumoto et al. (\cite{Mat01}) and Kiss et al.
(\cite{kiss}).

\noindent In this paper we deduce a detailed model of the system
basing on our optical photometry of 2001 and 2002. Further we
follow the final decline to quiescence and find quasi periodic
short timescale variations in this system. Together with
pre--outburst data of Honeycutt (\cite{honey01}) we finally
determine a period change.


\section{The Data}
The new data were obtained with the Innsbruck 60cm telescope
(Kimeswenger \cite{ag01}) and a direct imaging CCD device in the
period from June 21, 2001 to July 9, 2002. In 2001 a CompuScope
Kodak 0400 CCD (4\farcm6 $\times$ 3\farcm1 field of view) was
attached, in 2002 it was an AP7p SITe 502e (8\farcm36 $\times$
8\farcm36). 781 images were taken in 34 nights with $V$, $R$ and
$I_{\rm C}$ filters. The exposure times varied with filter,
brightness, weather conditions and camera. Flatfield and bias
subtraction were carried out in standard manner with the help of
MIDAS routines. The source extraction was performed using
SExtractor V2 (Bertin \& Arnouts \cite{sex}). The rms of the
comparison standards in the field was $<0\fm08$ with the Kodak
chip and $< 0\fm05$ with the SITe CCD typically.
The light curve was obtained by means of differential photometry
of up to 40 stars within about 4\farcm0 from the target in this
very crowded field. For the absolute calibration CCD standards in
the field by Henden \& Honeycutt (\cite{H_H}) and Henden
(\cite{Henden01}) were applied. We found on that occasion that the
coordinates of the whole set of the Henden \& Honeycutt standards
are shifted 1\farcm73 east and 1\farcm20 south and thus the
finding chart overlay (SIMBAD/ALADIN) is wrong. As we use Johnson
$R$ and the CCD standards of Henden (\cite{Henden01}) are taken in
$R_{\rm C}$ we have to assume color terms in the absolute
calibration. The attempt to derive these terms failed within the
accuracy of the photometric data. We thus did not use the R band
for the final results in extinction and distance (see
section~\ref{dist_s}).

The pre--outburst data ranging from June 4, 1991 to September 29,
1996 were provided to us by Honeycutt (\cite{honey01}). The data
consist of two sets (changing June 1995) with different zero
points. The first set (the one used for Mennickent \& Honeycutt
\cite{M_H}) was shifted according to the information given by
Honeycutt to overlay. This corresponds well to the calibration of
Skody \& Howell (\cite{skody}).

\section{The Photometric Behavior}
\label{pho} 
\begin{figure}[ht]
\centering
\includegraphics[width=85mm]{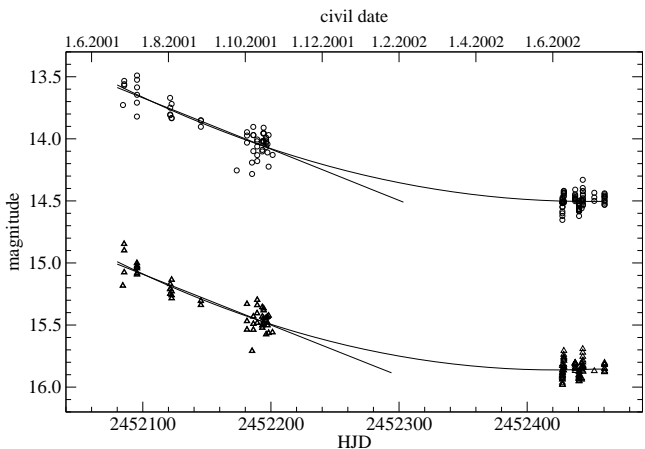}
\caption{The final decline phase of the recent nova outburst
(upper: $I_{\rm C}$; lower: $V$). Data around primary minima were
removed to decrease effects due to the orbital modulation. First
order polynomials to the 2001 data and second order polynomials to
the whole data set are shown.} \label{decline}
\end{figure}

First of all we derived the 'remnant' of the outburst
in our 2001 data. The object clearly had not returned completely
to quiescence as stated by Schaefer (\cite{ci_4}) for August 4,
2001. To verify the real level, the new period (see
section~\ref{sec_period}) was taken and the data points outside
the primary eclipse were used to derive the final decline of the
nova outburst. We see (Fig.~\ref{decline}) that this decline was
finished somewhen in February/March 2002.

\begin{figure}[ht]
\centering
\includegraphics[width=85mm]{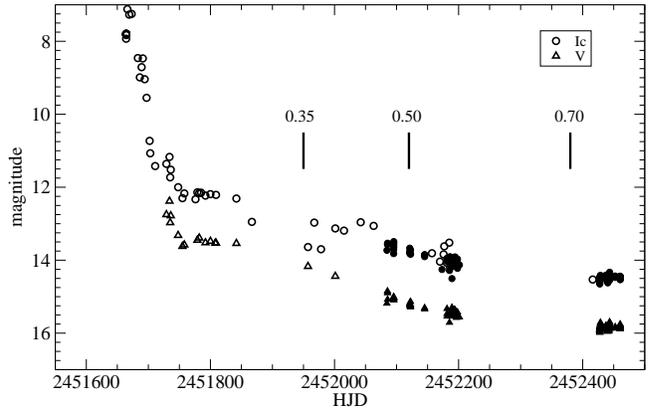}
\caption{The complete decline phase of the 2000 outburst (open
symbols: CCD data from VSNET; closed symbols: our photometry). Due
to the orbital modulation the upper boundaries have to be used.
The ticks mark the predicted dates for the abrupt decline after
the plateau for different hydrogen content $X$ (Hachisu \& Kato
\cite{HaKa}).} \label{total}
\end{figure}

In contrast to this slowly continuous decline, the plateau phase
in the model of Hachisu \& Kato (\cite{HaKa}) should stop abruptly
when the super--soft X--ray source phase ends. The position for
this end of the plateau phase depends strongly on the hydrogen
content $X$. The overall evolution is shown in Fig.~\ref{total}.
We find that the decline started between JD~2452050 and
JD~2452100.
\begin{figure}[h]
\centering
\includegraphics[width=65mm]{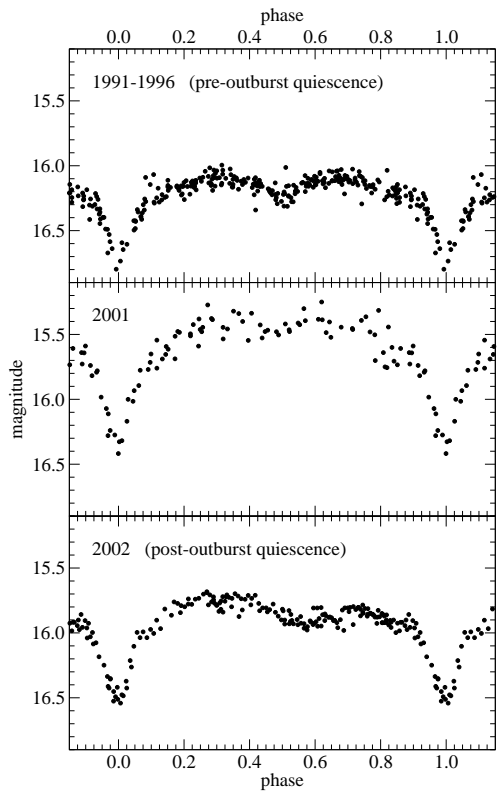}
\caption{The $V$ light curves for three different epochs:
1991-1996 (upper), 2001 (middle) and 2002 (lower panel). The light
curve for 2001 was corrected according to the  linear final
decline shown above and normalized relative to October 1, 2001.
The scatter is dominated by the short timescale variations (see
section~\ref{variations}) and not by the errors in the data.}
\label{lightcurve}
\end{figure}
This results, including the late decline model calculations of
Hachisu \& Kato (\cite{HaKa}), in a hydrogen content of
\mbox{$0.45 < X < 0.50$}. This is in contradiction to Hachisu \&
Kato (\cite{HaKaII}), who obtained a hydrogen poor model with
\mbox{$X = 0.35$} on the basis of only two data points from the
IAU Circulars of Schaefer (\cite{ci_4}, \cite{ci_5}). Our
moderately hydrogen rich result is supported by the infrared
spectroscopy of Lynch et al. (\cite{ci_6}) and by the first
decline model (Fig.~4 in Hachisu \& Kato \cite{HaKa}).

\noindent The linear decline (Fig.~\ref{decline}) was subtracted
to derive the light curve for 2001 which is shown together with
those from the pre--outburst and from 2002 in
Fig.~\ref{lightcurve}. The periods and epochs used are described
in section~\ref{sec_period}.

\section{The 3D Model}
The double star system was modelled with a slightly evolved main
sequence secondary (SE) filling the Roche lobe, a nearly point
source like, spherically symmetric white dwarf (WD), a
rotationally symmetric inner accretion disk, a thin accretion
stream from the Lagrange point towards the disk and a spray of
bounced material first introduced by Schandl et al.
(\cite{spray}).

The WD mass of \mbox{1.2\,\,M$_\odot$} was derived by Hachisu \&
Kato (\cite{HaKa}, Fig.~4) by means of the thick wind model of the
early decline of the 1917 and the 2000 outburst. The mass of the
SE can not be determined that directly. For a range of SE masses
we used the evolutionary tracks and colors for solar abundance
stars of Girardi et al. (\cite{tracks}) and our 2002 photometry at
minimum, when the SE dominates the emission. The point in the
evolutionary track was chosen to fill the Roche lobe. This results
in the extinction free ($V-I_{\rm C}$)$_0$ color. Together with
our measured ($V-I_{\rm C}$) in 2002 and each SE mass we are able
to derive a corresponding range for E$_{\rm B-V}$. The results are
summarized in Tab.~\ref{limit}.

\begin{table}[h]
\caption{Extinction derived from ($V-I_{\rm C}$)$_{minimum} =
1\fm5$ \protect\newline and the evolutionary tracks at Roche lobe
filling position.} \label{limit}
\begin{tabular}{c c c c c}
$M_{\rm SE}$ $[$M$_\odot]$ & $R^{\rm RL}_{\rm SE}$ $[$R$_\odot]$ & $T_{\rm eff}$ $[$K$]$& ($V-I_{\rm C}$)$_0$ & E$_{\rm B-V}$ \\
\hline
1.4 & 1.65 & 6\,690 & 0\fm488 & 0\fm82 \\
1.5 & 1.69 & 7\,040 & 0\fm405 & 0\fm90 \\
1.6 & 1.74 & 7\,380 & 0\fm307 & 0\fm97 \\
1.7 & 1.78 & 7\,760 & 0\fm214 & 1\fm04 \\
1.8 & 1.83 & 8\,110 & 0\fm132 & 1\fm10 \\
\end{tabular}
\end{table}

According to van den Heuvel et al. (\cite{vH}) a 1.4 M$_\odot$
ZAMS star as SE can already achieve a mass transfer, which leads
to a steady hydrogen shell burning on the WD. A recurrent nova is
not likely anymore below this mass limit. Thus a smaller SE does
not have to be taken into account. On the other hand the $VI_{\rm
C}JK_{\rm s}$ photometry of Schmeja et al. (\cite{schmeja})
shortly after the outburst limits the interstellar extinction. The
($V-K_{\rm s}$)$_0$ as well as the ($I_{\rm C}-J$)$_0$ were
calculated for different interstellar extinctions. They get to a
negative domain at \mbox{E$_{\rm B-V} > 1\fm05$}. This is thus an
upper limit for the extinction, assuming a dominant photosphere.
Thus we conclude \mbox{$1.4 < {M}_{\rm SE} \leq 1.7$ M$_\odot$}.

The geometric model was realized with the help of the "{\sl MATLAB
language of technical computing 5.3R11}" (\cite{matlab}). This
allowed us to easily implement and modify various asymmetric
surfaces, system inclinations and rotation periods. All surfaces
are assumed to radiate like black bodies. The ray tracing for the
irradiation within the system was solved explicitly. The ray
tracings during system rotation for the resulting light curves
were obtained by the internal renderer.
\begin{figure}[ht]
\centering
\includegraphics[width=85mm]{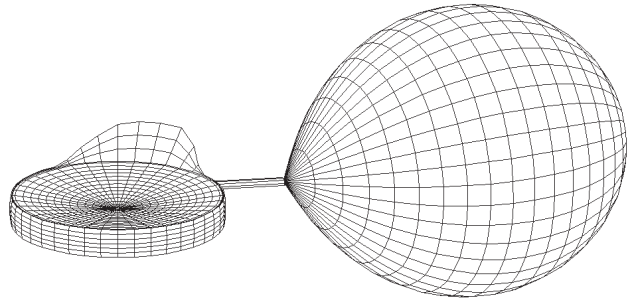}
\caption{The basic geometric model with the Roche lobe filling SE,
the symmetric accretion disk, the accretion stream and the spray
of bounced material at the light curve phase of 0.33. The grid
mesh density of the surfaces was reduced here for clarity  by a
factor of three with
 respect to the one used in the real calculations.} \label{model_1}
\end{figure}
The model consists of the five components described in detail
later. The resulting scheme is shown in Fig.~\ref{model_1}. Each
surface element was characterized by its temperature. To obtain
the flux in each band, the contribution of each element was
calculated by folding the black body with the filter curves ($B$,
$V$, $R_{\rm C}$ and $I_{\rm C}$: Bessel \cite{bessel}; Johnson
$R$: Aller et al. \cite{LB}). For the zero point a spherical body
simulating the sun was used in the same way in the renderer and
calibrated by the values of Bessel et al. (\cite{bessel2}).

\paragraph{The White Dwarf\newline}
We assume a spherical WD photosphere. The luminosity of the WD is
given by
\begin{equation}
L_{\scriptscriptstyle\rm WD} = L_{{\scriptscriptstyle\rm HSB}} +
{1\over2}\,{{\rm G}\,M_{\scriptscriptstyle\rm WD}\,\dot M\over
R_{\scriptscriptstyle\rm WD}}
\end{equation}
\noindent where $L_{{\scriptscriptstyle\rm HSB}}$ is the
luminosity obtained from the remnant of the hydrogen shell burning
in 2001 (Hachisu \& Kato \cite{HaKa}) and the second term
originates from the accretion (Shaviv \& Starrfield
\cite{star87}). According to the model of Hachisu \& Kato and the
photometric behavior shown before, there is a remnant of the
thermonuclear processes after the outburst in the 2001 data. The
simulations give us \mbox{$L_{{\scriptscriptstyle\rm HSB}} \approx
200\,$L$_\odot$}. This results in a net hydrogen mass burning
\mbox{$\dot M_{\rm He} \approx 10^{-9}$\,\,M$_\odot$} during the
plateau phase from end 2000 to August 2001. This is just below
{$^1/_{10}$} of the mass-increase rate given by Hachisu \& Kato
for the total outburst.

The direct light contribution of the WD is negligibly small in our
visual bands. It only contributes by the irradiation on the other
components.

\paragraph{The Secondary\newline}

\begin{figure}[ht]
\centering
\includegraphics[width=88mm]{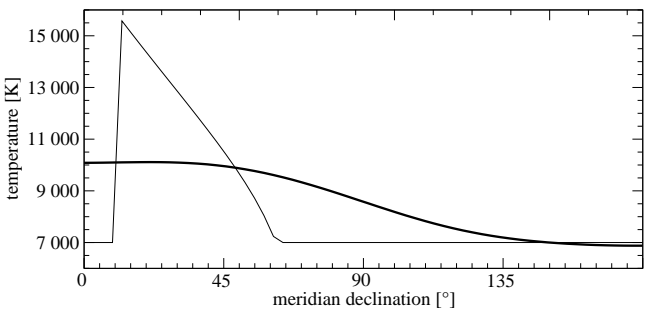}
\caption{The effects of the redistribution of the temperature
along the meridian from the inner critical Lagrange point (0\degr)
to a pole (90\degr) and further to the backside of the SE
(180\degr): thick line - redistributed temperature profile; thin
line - original irradiated temperature $T_{*,irr}$.}
\label{redistribute}
\end{figure}

The shape of the secondary is calculated numerically. It starts in
the inner critical Lagrange point which is fixed by
\begin{equation}
\left({2 \pi \over
P}\right)^2\left({\stackrel{\rightarrow}{r}_{\!\scriptscriptstyle\rm
CG,0} \cdot
\stackrel{\rightarrow}{e}_{\!\scriptscriptstyle1}}\right)
\,= 
{{\rm G}M_{\scriptscriptstyle\rm WD} \over
|{\stackrel{\rightarrow}{r}_{\!\scriptscriptstyle\rm
WD,0}}|^2}
- \frac{{\rm G}M_{\scriptscriptstyle\rm SE}}{
|{\stackrel{\rightarrow}{r}_{\!\scriptscriptstyle\rm
SE,0}}|^2}
\end{equation}
\noindent where G is the gravitational constant and
${\stackrel{\rightarrow}{r}_{\!\scriptscriptstyle\rm WD}}$,
${\stackrel{\rightarrow}{r}_{\!\scriptscriptstyle\rm SE}}$ and
${\stackrel{\rightarrow}{r}_{\!\scriptscriptstyle\rm CG}}$ the
vectors to the WD, the SE and the center of gravity (CG)
respectively. $\stackrel{\rightarrow}{e}_{\!\scriptscriptstyle1}$
is the unity vector parallel to
${\stackrel{\rightarrow}{r}_{\!\scriptscriptstyle\rm CG}} -
{\stackrel{\rightarrow}{r}_{\!\scriptscriptstyle\rm SE}}$. An
index 0 stands for values at the Lagrange point. The surface of
the SE then follows the equipotential $\phi = \phi_0$ including
the rotation of the complete binary flattening the system
\begin{equation}
\phi_0 = - {\rm G}\,\, \left({{M_{\scriptscriptstyle\rm WD} \over
|{\stackrel{\rightarrow}{r}_{{\!\scriptscriptstyle\rm WD}}}|} +
\frac{M_{\scriptscriptstyle\rm SE}}{
|{\stackrel{\rightarrow}{r}_{{\!\scriptscriptstyle\rm
SE}}}|}}\right) - \frac{\pi}{P}
\,\,\left({\stackrel{\rightarrow}{r}_{\!\scriptscriptstyle\rm
CG} \cdot
\stackrel{\rightarrow}{e}_{\!\scriptscriptstyle1}}\right)^2
\end{equation}
\noindent The calculated volume corresponds within one
percent to the Roche lobe volume of Eggleton (\cite{roche}). \\
The star is irradiated by the WD. This increases the undisturbed
temperature $T_{*,0}$:
\begin{equation}
\sigma T_{*,irr}^4 = \sigma T_{*,0}^4 +
\eta_*\,{L_{\scriptscriptstyle\rm WD}\over 4\pi r^2} \,\cos \theta
\label{e4}
\end{equation}
where $r$ is the distance of the surface element from the WD and
$\theta$ the angle of incidence. An efficiency $\eta_* = 0.7$ was
used. The value results from fitting details of the light curve -
mainly the gradient at phases 0.08 to 0.18. This somewhat higher
value than that assumed in previous investigations may originate
from the compact MS nature of the SE preventing the star from
using the irradiated energy to change its internal structure. For
the model this energy is distributed along the stellar surface
(Fig.~\ref{redistribute}) in the same way as described in Schandl
et al. (\cite{spray}).

\paragraph{The Disk\newline}
A rotational symmetric accretion disk was used. Elongated disks as
they are proposed by Paczy\'nski (\cite{rd_1}) or result from
projecting the 3D disk models of Hirose et al. (\cite{3ddisk})
were tested too but showed negligible effects. The horizontal size
$R_{\rm D} = 0.75 R^{\scriptscriptstyle\rm
RL}_{\scriptscriptstyle\rm WD}$ of the disk is determined by the
stability due to tidal forces near the Roche lobe (Paczy\'nsky
\cite{rd_1}, Papaloizou \& Pringle \cite{rd_2}). The width of the
primary minimum independently leads us to the same result. The
vertical height in the inner part ($r < R_1$) was obtained from
the fit to the hydrodynamic models (Meyer \& Meyer-Hofmeister
\cite{MeMe}) by Schandl et al. (\cite{spray}). The calculations of
Meyer \& Meyer-Hofmeister (\cite{MeMe}) show a rapid turnover of
the vertical behavior. The lower gradient causes for those regions
no irradiation by the WD anymore. Therefore we used
\begin{equation}
z = \left\{
\begin{array}{l@{\quad}l}
 2.527\,\, r^{1.093}
\,\,M_{\scriptscriptstyle\rm WD}^{-0.38} \, {\dot M}^{0.17} & :\quad r
\leq
R_1\\
 & \\
 const. & :\quad R_1 \leq r \leq R_{\rm D}
 \end{array}\right.
\end{equation}
\noindent where $z$, $r$, $R_1$, $M_{\scriptscriptstyle\rm WD}$
and ${\dot M}$ are in solar units. $R_1$ was estimated from Fig.~1
in Meyer \& Meyer-Hofmeister (\cite{MeMe}) for
\mbox{$M_{\scriptscriptstyle\rm WD} = 1.2$\,M$_\odot$} to be
\begin{equation}
R_1 = 0.365 \,\log(\dot M) + 2.69
\end{equation}
\noindent This description of the disk differs significantly from
the parabolic law used by Hachisu \& Kato (\cite{HaKa}). They
argue, referring to Schandl et al. (\cite{spray}), that they
simulate the "flare--up" of the rim. But they increase strongly
the reradiation of the WD flux on every part of the disk. This
effect gets drastically at the outer disk portions. Moreover the
high "flare--up" built by the spray, as shown below, does not
cover the whole rim but only about a quarter of it. Meyer \&
Meyer-Hofmeister (\cite{MeMe}) also tested the effects of the
irradiation on the vertical structure and found only marginal
effects for CVs.\\
The disk temperature was adopted from the
frictional heating and the surface irradiation as in Schandl et
al. (\cite{spray}) by
\begin{equation}
\sigma T_{\scriptscriptstyle\rm D}^4 = {3 \over 8\pi}\,\,{{\rm G}
M_{\scriptscriptstyle\rm WD}\,\dot M \over
r^3}+\eta_{\scriptscriptstyle\rm D}\,{L_{\scriptscriptstyle\rm WD}
\over 4\pi r^2}\,\cos \theta
\end{equation}
For $T_{\scriptscriptstyle\rm D} < T_{\scriptscriptstyle ST}$ the
stream temperature $T_{\scriptscriptstyle ST}$  (see description
of the accretion stream later in this section) of the infalling
material was assumed at the disk too. The efficiency parameter for
the irradiation was assumed to be $\eta_{D} = 0.5$, like in
Schandl et al. (\cite{spray}) and Hachisu \& Kato (\cite{HaKa}).
Our tests with the models show nearly no noticeable effects on the
composite light curve for slightly different $\eta_{D}$. The black
body approximation fits very well for the disk in the visible
spectral range as shown by recent full radiative transfer models
(AcDc) producing artificial spectra (Nagel et al. \cite{AcDc}).

\paragraph{The Spray\newline}

\begin{figure*}[ht]
\centering
\includegraphics[width=140mm]{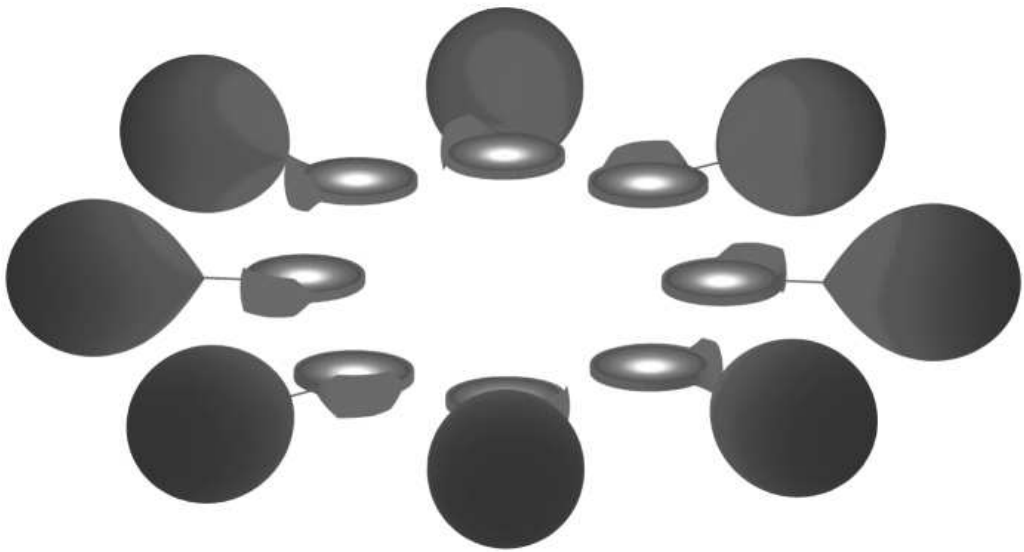}
 \caption{The model for the different phases. The grey scale represents
 the $V$ flux intensity due to the temperature distribution.} \label{model_3}
\end{figure*}

The spray was first introduced by Schandl et al. (\cite{spray}) as
convolute of individual blobs bounced after reaching the incident
point of the accretion stream. Schandl et al. (\cite{spray}) and
Meyer-Hofmeister et al. (\cite{sprays}, \cite{short_time}) found
that the vertical extension dominates the effect on the light
curve for the high inclination system here. Although they were
able to calculate individual trajectories of undisturbed blobs the
final, optically thick surface was modelled from the effects in
the light curve (see Fig.~\ref{model_3}).
\begin{figure}[ht]
\centering
\includegraphics[width=80mm]{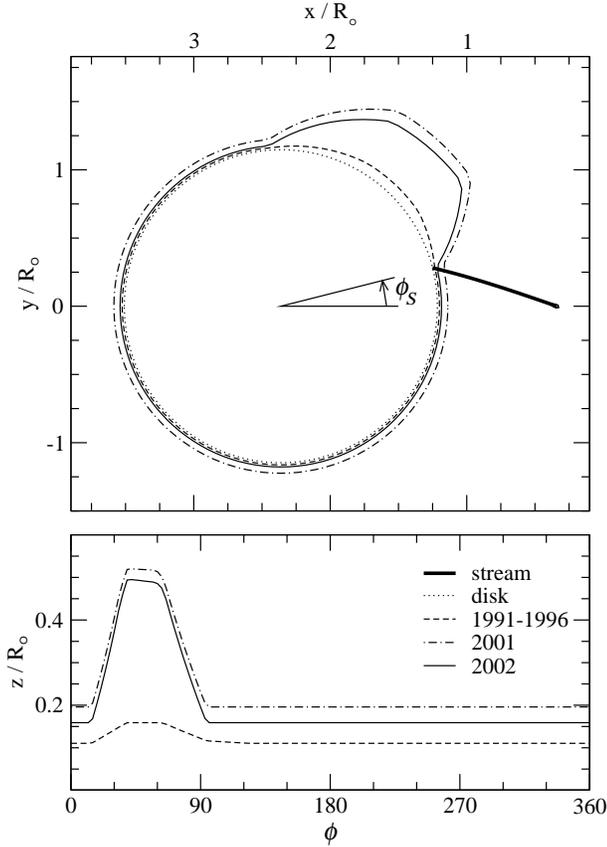}
\caption{The structure of the spray as top view (upper panel,
including the circular disk) and its vertical structure projection
(lower panel). The shape is symmetric with respect to the
equatorial plane. The coordinates originate in the center of
gravity (GC). For the stream the real trajectory is shown here.
For the calculation a straight approximation was used.}
\label{spray}
\end{figure}
We obtained the shape of the spray in the same way for our target
(Fig.~\ref{spray}). The massiveness of the spray varied with the
mass transfer rates.

\noindent The temperature of the irradiated regions of the spray
were deduced according to Eqn.~\ref{e4} with the stream
temperature $T_{\scriptscriptstyle ST}$ as undisturbed value.
Since the material consists of a clumpy medium the overall
efficiency for heating due to irradiation is not comparable to a
stellar photosphere. At short wavelength it is assumed that the
spray is even half transparent (Meyer \cite{meyer_private})
causing the observed super soft X-ray effects. We therefore found
in the simulations the plausible value of $\eta_S = 0.2$.

\paragraph{The Accretion Stream\newline}

The end points of the accretion stream are defined on the one hand
by the Lagrange point and on the other hand by the impact point.
The latter is calculated from an undisturbed gravitational
particle trajectory towards the disk and characterized by the
angle $\phi_{\scriptscriptstyle\rm S}$. As the bend of the
trajectory is small (Fig.~\ref{spray}), a straight line was
assumed. The cross--section of the stream is determined like in
Meyer \& Meyer-Hofmeister (\cite{MeMe83}) but slightly flattened.
Recent hydrodynamic numerical simulations by Oka et al.
(\cite{oka}) show very similar deflection angles and tube
geometries at $\tau \approx 1$.

\noindent The temperature $T_{\scriptscriptstyle ST}$ of the
material is defined by the irradiated and distributed temperature
of the SE at the stream source. The contribution of the stream to
the light curve is rather small.

\begin{figure*}[ht]
\sidecaption
 \includegraphics[width=108mm]{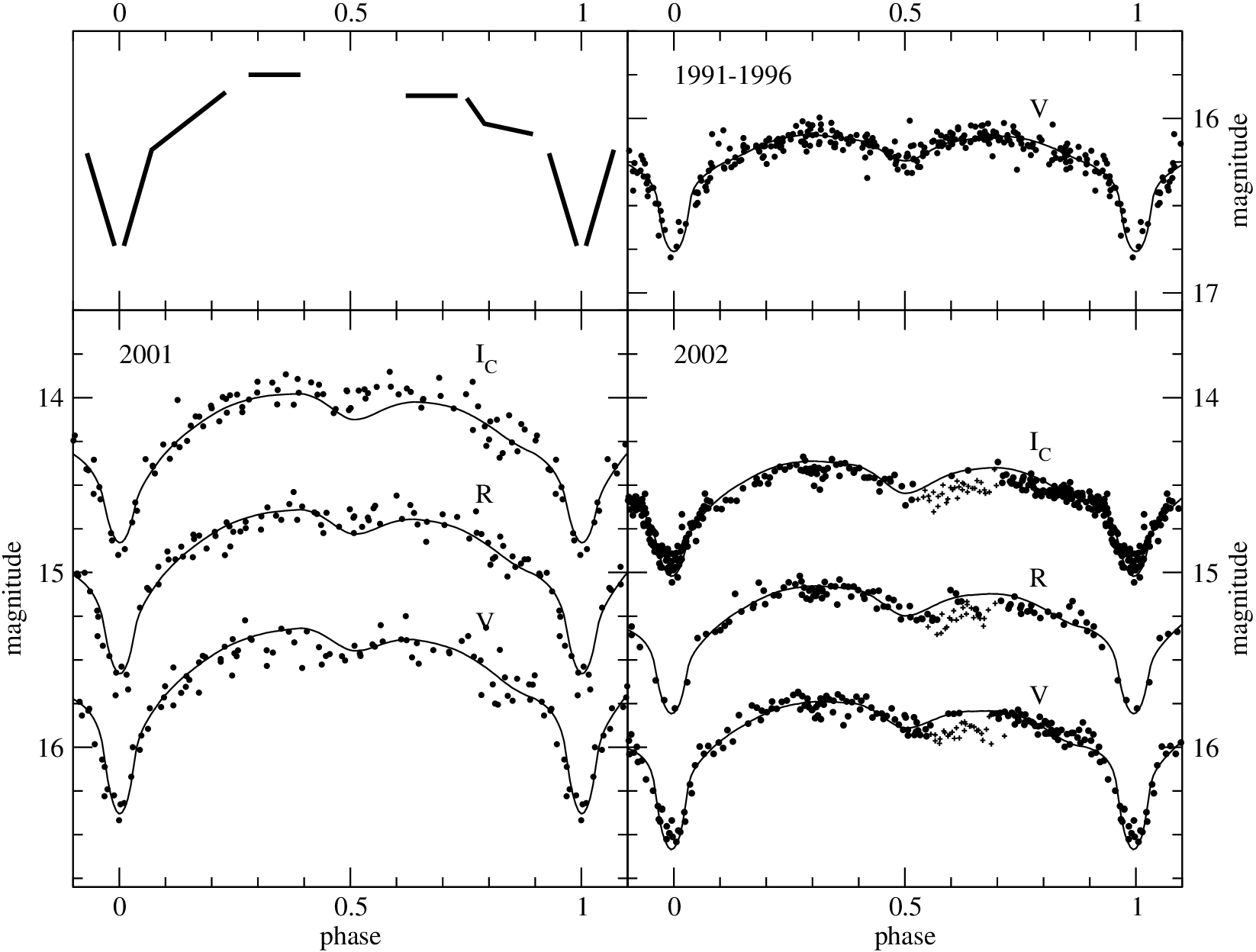}
\caption{~The features used to limit independently different
geometrical structures of the model (upper left panel). The data
points for the individual epochs and colors together with the
simulated curves are shown in the other panels. The crosses in the
2002 data from phase 0.55 to 0.68 correspond to data of two
individual nights only. They seem do deviate systematically from
the model. As shown later individual nights may vary due to
variations in the accretion rate. We assume this sub--sample to be
affected in such a way.} \label{model_2}
\end{figure*}

\paragraph{The Composition of the Modell\newline}

The individual components are dominating different parts of the
light curve:
 \begin{itemize}
 \item The width of the central part of the primary minimum gives
 the disk size and limits the system  inclination $i$.
 \item The depth of the primary and the secondary minimum strongly
 depend on the system inclination. Neither an inclination of
 70\degr\ nor of 72\degr\ result in a
usable fit, as in the first case the modelled primary minima are
not deep enough for reasonable $\dot M$ and adjusted secondary
minima, otherwise those of 2002 are too deep.
 \item The difference of the plateau levels before and after the
 secondary minimum gives the parameters of the secondary - namely
 the reduced irradiation by the shadow of the spray - and the
 irradiation efficiency of the spray.
 \item The depression of the late light curve gives length and
 height of the spray.
 \item The gradient of the late part of the primary minimum (phase: 0.08
 - 0.18)
 gives the irradiation parameter $\eta_{\scriptscriptstyle\rm SE}$ of the SE
\end{itemize}

\begin{figure}[ht]
\centering
\includegraphics[width=88mm]{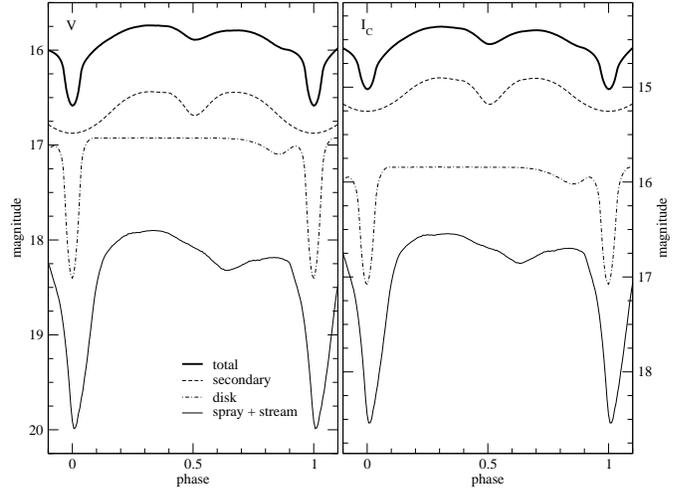}
 \caption{The contribution of the individual components to the total
 light curve in 2001 for $V$ (left) and $I_{\rm C}$ (right). The asymmetry
 of the spray causes a shift of the position in the primary eclipse. This
 is of special interest if combining data from different epochs
 (respectively mass transfer rates) for determination
 of the period.}
 \label{model_4}
\end{figure}

\phantom{X}

\begin{table}[ht]
\caption{The parameters of the best fit model.} \label{param}
\begin{tabular}{l l l}
system: &  i = 71$\degr$ \smallskip \\
secondary: & $M_{\scriptscriptstyle\rm SE}$  = 1.5 M$_\odot$ \smallskip & $T_{\scriptscriptstyle\rm SE}$ = 7\,000 K \\
white dwarf: & $M_{\scriptscriptstyle\rm WD}$  = 1.2 M$_\odot$ \smallskip & $R_{\scriptscriptstyle\rm WD}$ = 0.0072 R$_\odot$  \\
disk: & $R_{\scriptscriptstyle\rm D} = 0.75\,R^{\rm RL}_{\scriptscriptstyle\rm WD}$ \smallskip  \\
stream: & $\phi_{\scriptscriptstyle\rm S} = $14\fdg2 \smallskip &  \\
1991-1996 & $\dot {\rm M} = 2.5 \times 10^{-8}\,\,$M$_\odot$ & $R_1 = 0.54 \,R^{\rm RL}_{\scriptscriptstyle\rm WD}$\\
2001      & $\dot {\rm M} = 1.5 \times 10^{-7}$ & $R_1 = 0.75 $ \\
2002      & $\dot {\rm M} = 5.5 \times 10^{-8}$ & $R_1 = 0.72 $
\end{tabular}
\vspace{-3mm}
\end{table}
\relax

The parameters $M_{SE}$, $T_{SE}$, $i$, $\dot M$, $R_D$, $\eta_*$,
$\eta_D$, $\eta_S$ and the spray geometry were varied
independently in the beginning. The use of different photometric
bands -- namely $V$, $R$ and $I_{\rm C}$ -- gives us further
restrictions to the overlapping part of the possible individual
solutions. Finally $M_{SE}$, $T_{SE}$, $i$, $\eta_*$, $\eta_D$ and
$\eta_S$ are limiting the variation ranges between the years. The
pre--outburst parameters are missing other bands than $V$ and thus
are somewhat more flexible. The resulting simulated light curves
are shown together with the data in Fig.~\ref{model_2}. In
Fig.~\ref{model_4} the individual contributions of the components
to the total light curve are plotted.

\section{Distance and Interstellar Extinction}
\label{dist_s} 
\begin{figure}[ht]
\centering
\includegraphics[width=85mm]{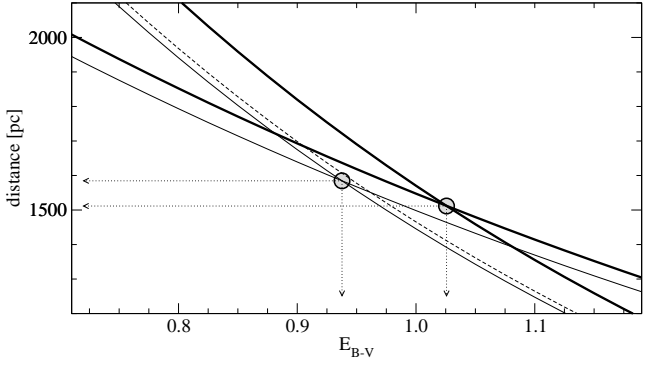}
\caption{The distance as function of the interstellar extinction
for different bands individually. The solutions are derived by the
intersection of the pairs of each year. The thick lines are the
solution for 2001 resulting in E$_{\rm B-V} = 1\fm025$. The thin
lines represent the 2002 data giving E$_{\rm B-V} = 0\fm935$. The
dashed line represents the pre--outburst data. Due to the lacking
second photometric band, no solution is possible there.
Nevertheless the tendency gives us some information on the
accuracy of the solutions above.} \label{dist}
\end{figure}

The model light curves were derived independently
for each band. When shifting those curves to the data we obtain
for each band an optimized solution as function of the distance
$D$ and the foreground extinction. Having more than two bands
allows us to derive  E$_{\rm B-V}$ (Fig.~\ref{dist}), using a
standard model for the interstellar extinction (Mathis et al.
\cite{MRN}).

Kiss et al. (\cite{kiss}) derive values of 0\fm83$\pm$0\fm20,
1\fm08$\pm$0\fm20 and 0\fm66$\pm$0\fm30 for the diffuse
interstellar bands (DIB) at 584.9~nm, 661.3~nm and CaII 393.4~nm
respectively. Using the photometries of Hanzl (\cite{hanzl}) and
Jesacher et al. (\cite{jesacher}) on days +2 and +6 after maximum
and the general tendency for novae around maximum of
\mbox{(B-V)$_0$ = 0\fm23$\pm0$\fm06} (Kiss et al. \cite{kiss}) we
get \mbox{$0\fm35 < $E$_{\rm B-V} < 0\fm75$}. But, as Kiss et al.
point out already, those colors may be affected strongly by
emission features. Hachisu \& Kato (\cite{HaKa}) derive from their
model of the pre--outburst light curve \mbox{E$_{\rm B-V}$ =
0\fm86}. In their revision (Hachisu \& Kato \cite{HaKaII}), based
on the early final $B$ band decline points of Schaefer
(\cite{ci_4}, \cite{ci_5}) and thus the low hydrogen content (see
section~\ref{pho}), they move it to \mbox{E$_{\rm B-V}$ = 1\fm00}.
While the direct spectroscopic methods give similar results, the
photometry suffers from the fact, that an average color for novae
is assumed. The comparison with the models using the $B$ band may
also suffer from the fact, that this band is mostly affected by
the non grey opacities. A black body, as used by the models, may
not work properly. We mostly rely on the direct spectroscopic
methods and the models here using the red bands and thus use
\mbox{E$_{\rm B-V} = 0\fm98\pm0\fm1$}.

The total luminosity also results from the simulations. This gives
us a distance of \mbox{1.52 kpc} to \mbox{1.58 kpc}. Hachisu \&
Kato (\cite{HaKa}) get 1.6~kpc from the maximum outburst
brightness and using \mbox{E$_{\rm B-V} = 0\fm85$}. Whereas they
derive 1.1~kpc in their revision (Hachisu \& Kato \cite{HaKaII}).


\section{Fast Mass Transfer Variations}
\label{variations} 
\begin{figure}[ht]
\centering
\includegraphics[width=87mm]{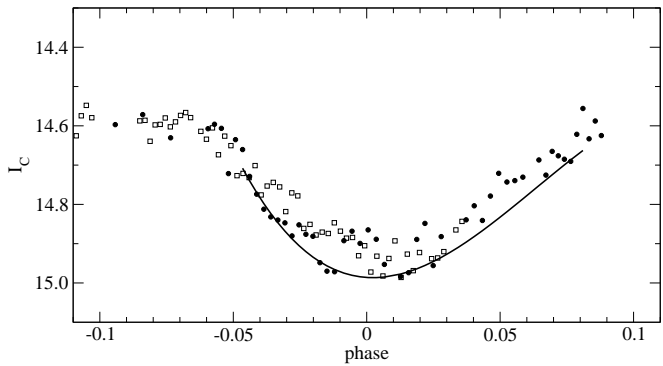}
\caption{The $I_{\rm C}$ minima of June 26, 2002 (circles) and
July 4, 2002 (squares). The line corresponds to the lower boundary
limit fit. } \label{minimum}
\end{figure}

The bandwidth of the photometric variations, as
shown in Fig.~\ref{lightcurve}, gives us information on short
timescale variations of the mass transfer. Meyer-Hofmeister et al.
(\cite{short_time}) found already in \mbox{RX~J0019.8+2156}
variations in the timescale of 1\fh2. To monitor the details of
the mass transfer we carried out fast $I_{\rm C}$ photometry in
the nights of June 26, 2002 and July 4, 2002. Both nights were
characterized by a primary minimum. The minima were overlayed and
the lower boundary was used to define the undisturbed minimum
(Fig.~\ref{minimum}).

\begin{figure}[ht]
\centering
\includegraphics[width=87mm]{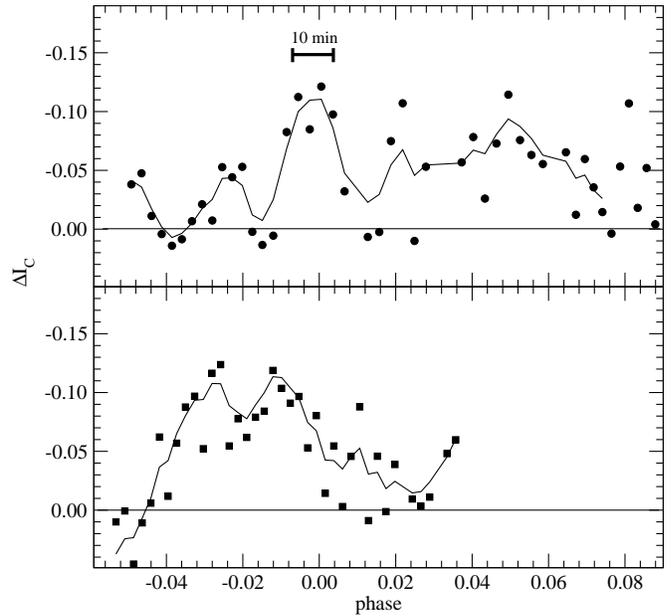}
\caption{The residuals of the $I_{\rm C}$ minima of June 26, 2002
(upper) and July 4, 2002 (lower). The lines are two period
floating means. The timescale of the small outbursts corresponds
to \mbox{$10^{\rm min} < t_{\rm F} < 16^{\rm min}$}.}
\label{min_res}
\end{figure}

After the substraction of the undisturbed minima, the residual
show the typical timescale of \mbox{$0.011 < t_{\rm F} < 0.018
\,\,P$} (\mbox{$10^{\rm min} < t_{\rm F} < 16^{\rm min}$}).
Meyer-Hofmeister et al. (\cite{short_time}) assume that some of
the individual blobs forming the spray are more violently expelled
due to variations and instabilities of the mass transfer flow.
This causes consequently a temporarily extended spray, detectable
as small "outburst". Meyer-Hofmeister et al. use
\mbox{${\pi\over2}\,t_{\rm K}$}, where $t_{\rm K}$ is the
Keplerian orbital period at the edge of the accretion disk. They
assume that the blobs oscillate free around the mean circular
orbit and thus give the timescale for the "outbursts". We, in
contrast, use the angular dimensions of the spray. The spray will
orbit with about the same rotational velocity as the accretion
disk. The geometric models result in a spray, where the expelled
material returns after about one quarter of an orbit to the
accretion disk. This results in \mbox{${1\over4}\,t_{\rm K}$} or
in our case 7$^{\rm min}$ ($0.008\,\,P$). As the blobs of the
spray reach out to up to 1.4 times the accretion disk radius the
timescale can be up to 12.5$^{\rm min}$ ($0.014\,\,P$). The finite
duration of the accretion outbursts extend those times. Thus the
calculated timescales are the lower boundaries. In fact we find
these variations during the minima, when the extended spray
geometry is well visible and at the same time the luminosity
contrast is preferring the effect in the photometry
(Fig.~\ref{min_res}).

\section{The Period - The Outburst Mass}
\label{sec_period} 
\begin{figure}[ht]
\centering
\includegraphics[width=87mm]{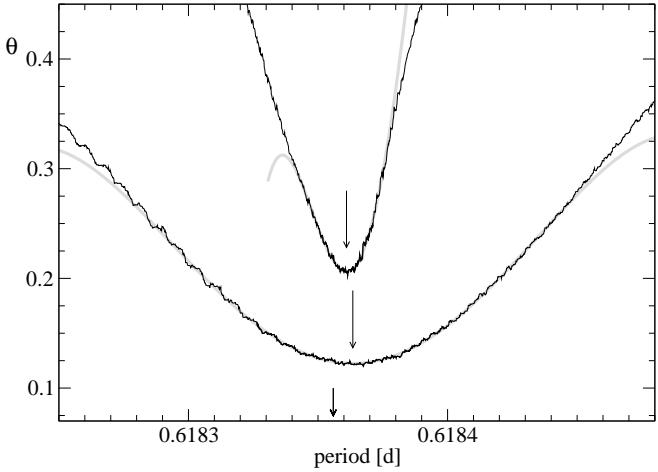}
\caption{The PDM $\theta$ cores of the $V$ data around the
minimum. The thick lines give the results from the data (upper:
Honneycutt \cite{honey01}; lower: our data). The grey lines are
polynomial fits of the minima cores to derive the exact positions.
The arrows mark the derived minima and the original period derived
by Mennickent \& Honeycutt (\cite{M_H}). } \label{pvm}
\end{figure}

\begin{figure}[ht] \centering
\includegraphics[width=87mm]{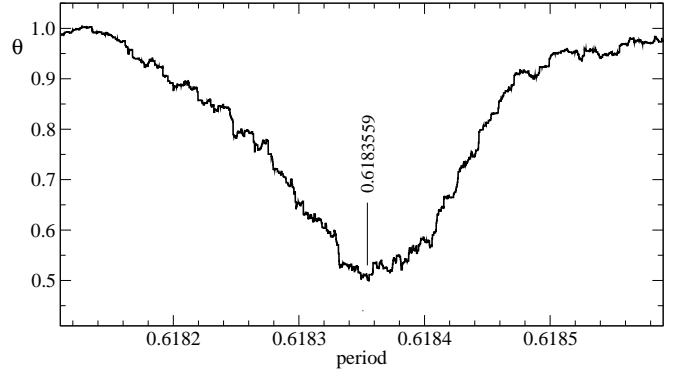}
\caption{The PDM diagram of the original data set of Mennickent \&
Honeycutt (\cite{M_H}) covering 1991 to April 1995 only by using
the default parameters for the binning by Stellingwerf
(\cite{pvm}). The original result is indicated at the position of
the lowest noise peak.} \label{diff_Honey_minima}
\end{figure}

Our data pointed out a shift of the period. The
periods were calculated by using  the PDM method (Stellingwerf
\cite{pvm}). The cores of the $\theta$ minima (Fig.~\ref{pvm})
were fitted by polynomials to define the exact minima. To obtain
homogeneity, the data by Honeycutt (\cite{honey01}) were used to
recalculate the pre--outburst parameters. To demonstrate the
necessity of this recalculation
\mbox{Fig.~\ref{diff_Honey_minima}} shows the PDM minima,
containing only the first data set by Mennickent \& Honeycutt with
the default sampling parameters of Stellingwerf (\cite{pvm}). The
absolutely lowest noise peak leads to the original result.
Recalculation with optimized parameters for PDM shows different
results. We thus propose the light curve elements of
$$ 2448412.147(46)\mbox{~} + 0\fd6183609(9) \times E$$
for the pre--outburst phase. For the post--outburst phase we get
$$ 2452081.5022(46) + 0\fd6183634(3)\times E.$$
Note that the epoch for the minima will slightly change with
decreasing accretion rate (Fig.~\ref{model_4}).


\noindent The period above is supported independently by the
$I_{\rm C}$ data giving us a period of 0\fd6183627 ($= 0.06$
seconds less). Using the change of the period $P$
\medskip
\begin{equation}
P^2 = {{4\pi^2a^3} \over {\rm G}\,M}; \qquad (P+\Delta P)^2 =
{{4\pi^2\hat a^3} \over {\rm G}\,(M-\Delta M)}
\end{equation}
\medskip
\noindent where $a$ and $\hat a$ are the separations before and
after the outburst, $M$ the total mass of the system, $\Delta P$
the period change and $\Delta M$ the outburst mass. The
conservation of angular momentum gives

\relax \medskip

\begin{equation}
\begin{array}{l}
\displaystyle{ {a^2 \over P}\,\,\left[{{M_{\scriptscriptstyle\rm
WD} \over \left({{M_{\scriptscriptstyle\rm WD} \over
M_{\scriptscriptstyle\rm SE}} + 1 }\right)^2}+
{M_{\scriptscriptstyle\rm SE} \over
\left({{M_{\scriptscriptstyle\rm SE} \over
M_{\scriptscriptstyle\rm WD}} + 1 }\right)^2}}\right]
= }\\
\\
\displaystyle{\xi \,{a^2 \over P}\,\, \left[{{\Delta M \over
\left({{M_{\scriptscriptstyle\rm WD} \over
M_{\scriptscriptstyle\rm SE}} + 1 }\right)^2}}\right] +}\\
\\
\displaystyle{{\hat a^2 \over P+\Delta
P}\,\,\left[{{M_{\scriptscriptstyle\rm WD} - \Delta M\over
\left({{M_{\scriptscriptstyle\rm WD}-\Delta M \over
M_{\scriptscriptstyle\rm SE}} + 1 }\right)^2}+
{M_{\scriptscriptstyle\rm SE} \over
\left({{M_{\scriptscriptstyle\rm SE} \over
M_{\scriptscriptstyle\rm WD} - \Delta M} + 1 }\right)^2}}\right]}
\end{array}
\end{equation}

\noindent The parameter $\xi \in [0,1]$ refers to the efficiency
of the angular momentum transfer to the expelled mass. The two
extreme are that the expelled mass carries away its original
angular momentum ($\xi = 1$) and the opposite case that all
angular momentum is transferred and thus kept in the system due to
viscosity ($\xi = 0$). We use these two cases for the calculated
range of mass loss given below. The resulting expelled mass does
not depend on the period $P$ within the tested range (100 times
the formal error). Also the mass of the secondary (testing
\mbox{$1.2 < $M$_{\scriptscriptstyle\rm SE} < 2.5 \,\,$M$_\odot$})
does not affect the result. The mass of the WD scales direct
proportional to $\Delta M$. As it is known to a 10-15\% level, it
does not influence $\Delta M$ significantly. The main
uncertainties originate from the errors of $\Delta P$ and from the
choice of $\xi$. Within a factor of 10 around the determined value
we got $\Delta P \propto \Delta M$. This gives us, using the
extreme values for $\xi$ and three times the formal error of
$\Delta P$
\[ \qquad 2.2 \times
10^{-6}\,\,{\rm M}_\odot\,\,\leq\,\,\Delta M\,\,\leq\,\,5.7 \times
10^{-6}\,\,{\rm M}_\odot.\] Hachisu \& Kato (\cite{HaKa}) obtained
\mbox{$4.7 \times 10^{-6}\,\,{\rm M}_\odot$} from the thick wind
model of the outburst. Assuming the last outburst 60 year ago, our
result corresponds to \mbox{$3.6\times 10^{-8}\,\,{\rm
M}_\odot/{\rm yr}\,\,\leq\,\,\dot {\rm M}\,\,\leq\,\,9.5 \times
10^{-8}\,\,{\rm M}_\odot/{\rm yr}$}. This corresponds well to the
accretion rates of the models during the quiescence phase.

\section{Conclusions}
The models presented allow us to determine the physical parameters
of the system within a very small range. This is mainly achieved
by the combination of data from different bands and different
epochs what is the main advantage to the work of Hachisu \& Kato
(\cite{HaKa}, \cite{HaKaII}). The model of the recurrent nova two
years after the outburst, when it reached a mean quiescent
photometric state still differs significantly from the one of the
pre--outburst phase.
Even speculative, one may assume that the energy transferred
during the outburst, when the secondary was completely enclosed by
the WD shell, caused a significant extension over the Roche lobe.
This increased the mass transfer to \mbox{$5.5 \times 10^{-8} <
\dot M < 1.5 \times 10^{-7}\,\,{\rm M}_\odot\,{\rm yr}^{-1}$} in
2001/2002. Thus if the mass transfer rate derived from the
pre--outburst phase is used as average, it does not reach up to
the accreted mass during the 60 years of quiescence. The mass
transfer rate of $\dot M \approx 2.5 \times 10^{-8}\,\,{\rm
M}_\odot\,{\rm yr}^{-1}$ derived for 1991-1996 differs from that
given by Hachisu \& Kato by a factor of four due to the different
model of the accreting region. The period shift determined in this
paper gives us a good estimate of the expelled mass $2.2 \times
10^{-6} < \Delta M < 5.7 \times 10^{-6}\,\,{\rm M}_\odot$. Within
the errors (the uncertainty due to the angular momentum carried by
the ejected material and the uncertainty of the average mass
transfer throughout the 60 years of quiescence) the determined
expelled mass corresponds to the accretion. Thus the evolution
towards the critical mass of the WD runs rather slow at this
evolutionary stage. Assuming a net mass increase of $<
10^{-6}$\,M$_\odot$ per outburst leads to a few 10$^7$ years to
reach the critical mass. On the other hand the evolutionary tracks
of the secondary show it at a position with extremely rapid
increase of its diameter (about
\mbox{5\,\,10$^{-10}$\,R$_\odot$~yr$^{-1}$}). Moreover the
decrease of the orbit is about 10$^{-7}$\,R$_\odot$ per outburst.
Thus the mass transfer should increase and may even evolve towards
steady hydrogen burning (van den Heuvel et al. \cite{vH}). The
derived distance and interstellar extinction gives us a somewhat
higher luminosity than in the outburst models of Hachisu \& Kato
(\cite{HaKa}). This affects also the considerations in Oka et al.
(\cite{oka}). Because of our  smaller inclination we also see some
parts of the disk during the primary minimum and thus it is not
necessary to increase the temperature of the SE in 2001 to obtain
a higher flux.

\begin{acknowledgements}
We like to thank R.K. Honeycutt and A.A. Henden for providing us
with their original measurements from the 1990's and the standards
in the field. We also thank F. Meyer (Munich) and P. Hauschild
(Georgia) for fruitful discussions. We are grateful to the VSNET
members for the data of the early decline.
\end{acknowledgements}


\begin{thebibliography}{}

\bibitem[1982]{LB}
Aller L.H., Appenzeller I., Baschek B., et al., 1982,
Landolt-B\"ornstein: Numerical Data and Functional Relationships
in Science and Technology - New Series Volume 2b, Springer Verlag
(Heidelberg)

\bibitem[1996]{sex}
Bertin E., \& Arnouts S., 1996, A\&AS, 117, 393

\bibitem[1990]{bessel}
Bessel M.S., 1990, PASP, 102, 1181

\bibitem[1998]{bessel2}
Bessel M.S., Castelli F., \& Plez B., 1998, A\&A, 333, 231

\bibitem[1983]{roche}
Eggleton P.P., 1983, ApJ, 268, 368

\bibitem[2000]{tracks}
Girardi L., Bressan  A., Bertelli G., \& Chiosi C., 2000, A\&AS,
141, 371

\bibitem[2001]{HaKa}
Hachisu I., \& Kato M., 2001, \apj, 553, L161

\bibitem[2002]{HaKaII}
Hachisu I., \& Kato M., 2002, in ASP Conf. Ser. 261, The Physics
of Cataclysmic Variables and Related Objects, ed. B.T. G\"ansicke
K. Beuermann, \& K. Reinsch, 627

\bibitem[2000]{hanzl}
Hanzl D., 2000, IAUC, 7444, 3

\bibitem[2001]{Henden01}
Henden A.A.,  2001, (8.11.2001) \newline
ftp.nofs.navy.mil/pub/outgoing/aah/sequence/ciaql.dat

\bibitem[1995]{H_H}
Henden A.A., \& Honeycutt R.K., 1995, PASP, 107, 324

\bibitem[1992]{vH}
Heuvel van den E.P.J., Bhattacharya D., Nomoto K., \& Rappaport
S.A., 1992, A\&A, 262, 97

\bibitem[1991]{3ddisk}
Hirose M., Osaki J., \& Mineshige S., 1991, PASJ, 43, 809

\bibitem[2001]{honey01}
Honeycutt R.K., 2001, private communication


\bibitem[2000]{jesacher}
Jesacher M.O., Kautsch S.J., Kimeswenger S., M\"uhlbacher M.S.,
Saurer W., Schmeja S., \& Scholz C.K., 2000, IAUC, 7426, 3

\bibitem[2001]{ag01}
Kimeswenger S., 2001, AG Abstr. Ser., 18, 251P

\bibitem[2001]{kiss}
Kiss L.L., Thomson J.R., Ogloza W., F\"ur\'esz G., \& Szil\'adi
K., 2001, A\&A, 266, 858


\bibitem[2002]{iauc7789}
Liller W., 2002, IAUC, 7789, 1

\bibitem[2002]{ci_6}
Lynch D.K., Wilson J.C., Miller N.A., Rudy R.J., Venturini C.C.,
Mazuk S., \& Puetter R.C., 2002, BAAS, 200, 75.05

\bibitem[1977]{MRN}
Mathis J.S., Rumpl W., \& Nordsieck K.H., 1977, ApJ, 217, 425

\bibitem[1999]{matlab}
MATLAB, 1999, The language of technical computing, MathWorks Inc.
(Natick, Mass.)


\bibitem[2001]{Mat01}
Matsumoto K., Uemura M., Kato T., et al. 2001, A\&A, 378, 487

\bibitem[1995]{M_H}
Mennickent R.E., \& Honeycutt R.K., 1995, IBVS, 4232, 1

\bibitem[2002]{meyer_private}
Meyer F., 2002, private communication

\bibitem[1982]{MeMe}
Meyer F., \& Meyer-Hofmeister E., 1982, A\&A, 106, 34

\bibitem[1983]{MeMe83}
Meyer F., \& Meyer-Hofmeister E., 1983, A\&A, 121, 29

\bibitem[1998]{short_time}
Meyer-Hofmeister E., Schandl S., Deufel B., Barwig H., \&
Meyer\,F., 1998, A\&A, 331, 612

\bibitem[1997]{sprays}
Meyer-Hofmeister E., Schandl S., \& Meyer F., 1997, A\&A, 321, 245


\bibitem[2002]{AcDc}
Nagel T., Dreizler S., \& Werner K., 2002, in ASP Conf. Ser.,
Workshop on Stellar Atmosphere Modeling, ed. I. Hubeny, D.
Mihalas, \& K. Werner., in press



\bibitem[2002]{oka}
Oka K., Nagae T., Matsuda T., Fujiwara H., \& Boffin H.M.J., 2002,
A\&A, in press, astro-ph/0208200

\bibitem[1977]{rd_1} Paczy\'nsky B., 1977, ApJ, 216, 822

\bibitem[1977]{rd_2} Papaloizou J., \& Pringle J.E., 1977, MNRAS, 181, 441


\bibitem[1925]{ci_1}
Reinmuth K., 1925, AN, 225, 385

\bibitem[2001a]{ci_3}
Schaefer B.E., 2001a, IAUC, 7750, 2

\bibitem[2001b]{ci_4}
Schaefer B.E., 2001b, IAUC, 7687, 5

\bibitem[2001c]{ci_5}
Schaefer B.E., 2001c, IAUC, 7621, 2

\bibitem[1997]{spray}
Schandl S., Meyer-Hofmeister E., \& Meyer F., 1997, A\&A, 318, 73

\bibitem[2000]{schmeja}
Schmeja S., Armsdorfer B., \& Kimeswenger S., 2000, IBVS, 4957

\bibitem[1995]{Sec}
Sekiguchi K., 1995, Ap\&SS, 230, 75

\bibitem[1992]{skody}
Skody P., \& Howell S.B., 1992, ApJS, 78, 537


\bibitem[1987]{star87}
Shaviv G., \& Starrfield S., 1987, ApJ, 321, L51

\bibitem[1978]{pvm}
Stellingwerf R.F., 1978, ApJ, 224, 953


\bibitem[1987]{We87}
Webbink R.F., Livio M., Truran J.W., \& Orio M., 1987, ApJ, 314,
653


\bibitem[2000]{ci_2}
Williams D.B., 2000, IBVS, 4904, 1

\end{thebibliography}
\end{document}